\documentclass{kluwer}    

\newdisplay{guess}{Conjecture}

\begin{document}                                                                                   
\begin{article}

\begin{opening}         
\title{Gamma-Ray Emission From Be/X-Ray Binaries}
\author{Mariana \surname{Orellana}}
\author{Gustavo E.\surname{Romero}\thanks{Member of CONICET. Email: {\sl romero@irma.iar.unlp.edu.ar}}}
\runningauthor{M. Orellana \& G.E. Romero}
\runningtitle{Gamma-Ray Emission From Be/X-Ray Binaries}

\institute{Instituto Argentino de Radioastronom\'{\i}a, C.C. 5, 1894 Villa Elisa, Argentina, {\rm and} \\Facultad de Ciencias Astron\'omicas y Geof\'{\i}sicas, UNLP, Paseo del Bosque, 1900 La Plata, Argentina.}

\date{June 1, 2004}

\begin{abstract}
Be/X-ray binaries are systems formed by a massive Be star and a magnetized neutron star, usually in an eccentric orbit. The Be star has strong equatorial winds occasionally forming a circumstellar disk. When the neutron star intersects the disk the accretion rate dramatically increases and a transient accretion disk can be formed around the compact object. This disk can last longer than a single orbit in the case of major outbursts. If the disk rotates faster than the neutron star, the Cheng-Ruderman mechanism can produce a current of relativistic protons that would impact onto the disk surface, producing gamma-rays from neutral pion decays and initiating electromagnetic cascades inside the disk. In this paper we present calculations of the evolution of the disk parameters during both major and minor X-ray events, and we discuss the generation of gamma-ray emission at different energies within a variety of models that include both screened and unscreened disks. 

\end{abstract}


\end{opening}           

\section{Introduction}

A significant number of variable gamma-ray sources of likely galactic origin has been found by satellite-borne experiments such as COMPTEL and EGRET (e.g. Torres et al. 2001, Zhang et al. 2002, Nolan et al. 2003). These sources vary their flux on timescales from weeks to months, similar to the timescales exhibited by X-ray transients at lower energies. The fact that some Be/X-ray binaries with major X-ray outbursts are located within the positional error boxes 
of unidentified and variable gamma-ray sources (e.g. Zhang et al. 2002; Romero et al. 2001, 2004) posits the problem of whether such systems are capable of generating non-thermal high-energy emission. In this paper we discuss how MeV, GeV and even TeV gamma-rays might be produced by accreting neutron stars in some Be/X-ray binaries and what is the expected time evolution of the high-energy radiation.

\section{Be/X-ray binaries} 

Be/X-ray binaries are the most abundant type of massive X-ray binary systems in the Galaxy, with an inferred population in the range $10^3-10^4$. These systems are formed by a rapidly rotating Be star and a neutron star, which is usually observed in an eccentric orbit. In some cases, known as Transient X-Ray Sources (TXRS), significant long-term X-ray variability is observed. These sources can vary from 1 to 3 orders of magnitude going from a quiescent phase to a major outburst, over timescales that can go from days to a few years. Sometimes, a major outbursts is followed by a series of smaller bursts. Perhaps the most extensively studied source of this class is A0535+26 (e.g. Giovanelli \& Sabau Graziati 1992), although a few tens are known.  
  
\begin{figure} 
\vskip 2.7in
\includegraphics{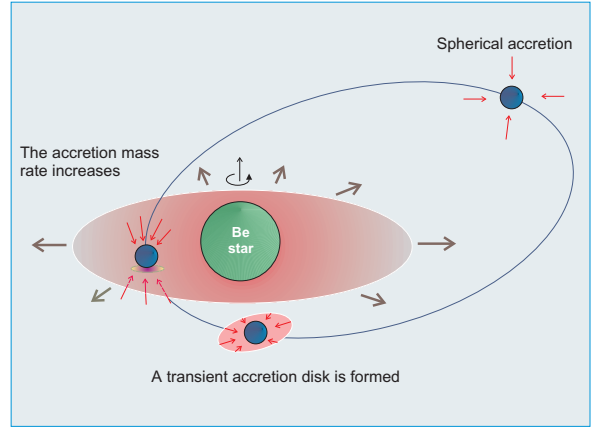}
\caption[]{Sketch of a Be/X-ray binary system}
\label{Sketch1}
\end{figure}

The general situation in a TXRS is depicted in Figure \ref{Sketch1}. The Be star loses mass in the form of a thick equatorial disk with a low expansion velocity and a low-density, fast polar wind. Far from the Be star, the neutron star accretes in the Bondi-Hoyle regime. Usually, X-ray flares occur once per orbit when the accretion rate increases dramatically close to the periastron passage. Occasionally, the equatorial disk of the primary can extend beyond the orbit of the compact star, which can then penetrate the disk leading to a major outburst. A transient accretion disk can be formed around the neutron star in such systems. The disk can survive during more than a single orbit in the case of a major outburst or it can last only for a fraction of the orbit in the case of minor outbursts. Evidence for the existence of such transient disks has been found through the detection of quasi-periodic oscillations (QPOs) in the power spectra of the hard X-ray flux of some objects (e.g. Finger et al. 1996a).

When the equatorial disk of the Be star retracts a major outburst may be followed by a series of minor ones in the same object, as observed, for instance, in the system 2S 1417-64 (Finger et al. 1996b).

\section{The magnetosphere of accreting neutron stars}

Once formed, the accretion disk can penetrate the magnetosphere of the rotating neutron star (Gosh \& Lamb 1979). This penetration creates a broad transition region between the unperturbed disk flow and the co-rotating magnetospheric flow. In this transition zone, at the inner radius $R_0$ the angular velocity is still Keplerian. Between $R_0$ and the co-rotation radius $R_{\rm co}$ there is a thin boundary layer where the angular velocity departs from the Keplerian value. At $R_{\rm co}$ the disk is disrupted by the magnetic pressure and the matter is channeled by the magnetic field lines to the neutron star polar cap, where it impacts producing hard X-ray emission.

\begin{figure} 
\vskip 2.7in
\includegraphics{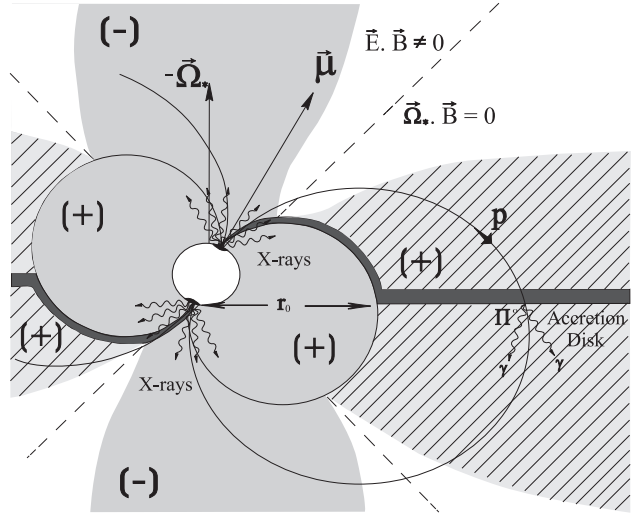}
\caption[]{Sketch of the magnetosphere of an accreting neutron star when $\Omega_*< \Omega_{\rm disk}$. From Romero et al. (2001).}
\label{Sketch2}
\end{figure}

Cheng \& Ruderman (1989, 1991) have studied the magnetosphere of accreting neutron stars when the disk rotates faster than the star, showing that three definite regions are formed: 1. A region coupled to the star by the magnetic field lines that do not penetrate the disk, 2. An equatorial region linked to the disk by the field attached to it, and 3. A gap entirely empty of plasma separating both regions. In this gap ${\bf E\cdot B}\neq 0$ and a strong potential drop $\Delta V$ is established (see Fig. \ref{Sketch2} for a sketch of the situation). For typical parameters in Be/X-ray binaries, $\Delta V\sim 10^{13-14}$ V (Romero et al. 2001). Protons entering into the gap from the stellar co-rotating region will be accelerated to multi-TeV energies and directed to the accretion disk by the field lines, where they will impact producing $\pi^{\pm}$ and $\pi^0$ in $pp$ interactions with the disk material.  These pions will decay inside the disk producing electron-positron pairs, high-energy (TeV) photons, and neutrinos. The neutrinos will escape from the system (see Anchordoqui et al. 2003 for estimates of the neutrino flux), but the pairs and the gamma-rays initiate electromagnetic cascades that will degrade the energy of the gamma-ray photons to form a standard cascade spectrum with a cut-off at MeV-GeV energies. The leptons in the disk will lose energy by inverse Compton (IC) interactions with the thermal X-ray photons, by synchrotron radiation in the field that penetrates the disk, and by relativistic Bremsstrahlung in the ions of the accreting gas. The details of the cascade development will be determined by the relative ratios of the cooling times among all these processes and the opacities for the produced photons in the specific cases.    

\section{Gamma-ray production in the accretion disk}

Protons entering into the gap from the stellar co-rotating region
are accelerated up to energies $E_p\sim  e\,V_{\rm max}$, where the potential drop $V_{\rm max}$ is given by (Cheng \& Ruderman 1991):
\begin{equation}
\!\!V_{\rm max} \sim 4 \times 10^{14}
\beta^{-5/2}\left(\frac{M_{*}}{M_{\odot}}\right)^{1/7}\!\!
R_6^{-4/7}L_{37}^{5/7}B_{12}^{-3/7}\;\;{\rm V}.
\end{equation}
Here, $M_{*} \sim 1.4$ $M_{\odot}$ is the mass of the neutron star, $R_6$
is its radius in units of $10^6$ cm, $B_{12}$ is the magnetic
field in units of $10^{12}$ G, and $L_{37}$ is the X-ray
luminosity in units of $10^{37}$ erg s$^{-1}$. The parameter
$\beta\equiv 2r_0/r_{\rm A} \sim 1$ is twice the ratio of the
inner accretion disk radius to the Alfv\'en radius (Gosh \& Lamb 1979). The Alfv\'en radius for spherical accretion can
be determined by the condition that the unscreened magnetic energy
density of the stellar field, $B^2/8\pi$, becomes comparable to
the kinetic energy density of the accreting matter, $\rho v^2/2$,
i.e. $ B(r_{\rm A})^2/8\,\pi= \rho (r_{\rm A})\,\, v(r_{\rm
A})^2/2\,\,, $ which yields $r_{\rm A}=\mu^{4/7} \,\dot{M}^{-2/7}
\,(2\,G\,M_*)^{-1/7}$, with $\mu$ being the dipolar magnetic
moment and $\dot{M}$ the accretion rate.

The ultra-relativistic proton flux will be directed to the accretion disk by the field lines.
The maximum current that can flow through the gap can be
determined from the requirement that the azimuthal magnetic field
induced by the current does not exceed that of the initial field
${\bf B}$:  $J_{\rm max}\sim B_{*}R_{*}^3 r_0^{-2}$ (Cheng \& Ruderman 1989); equivalently
\begin{equation}
 J_{\rm max}
\sim 1.5\times10^{24}
\beta^{-2}\left(\frac{M_*}{M_{\odot}}\right)^{- 2/7}
\!\!\!\!\!R_6^{1/7}L_{37}^{4/7}B_{12}^{-1/7}\;{\rm esu\;s^{-1}}.
\end{equation}

The number of protons that impact on the disk per unit of time is huge: $N_p=J_{\rm max}/e\sim$ $\cal{O}$ $(10^{33})$ s$^{-1}$ and the total power injected in the disk results: $\dot{E}_{\rm inj}\sim J_{\rm max}V_{\rm max}\sim$ $\cal{O}$ $(10^{36})$ erg s$^{-1}$.

Since all protons arrive to the disk surface with basically the same energy, we can estimate the proton flux as:
\begin{equation}
J_p (E) \sim \frac{\delta(E-E_p)}{2\pi}\frac{J_{\rm max}}{\pi e R_0^2}\;\;\; \frac{\rm protons}{\rm s \;cm^2\; sr}. \label{J_p}
\end{equation}
These protons will have a probability of penetrating up to a distance $L$ in the disk given by $P(L)=1-e^{-L/\lambda_{pp}}$, with $\lambda_{pp}=(\sigma_{pp} n^{\rm disk}_p)^{-1}$. At the relevant energies $\sigma_{pp}\sim 4\times 10^{-26}$ cm$^2$. Once the protons interact, neutral and charged pions are produced. The decay of the charged pions leads to neutrino production in the disk (Cheng et al. 1990, Anchordoqui et al. 2003) and the injection of new $e^{\pm}$-pairs. The gamma-ray emissivity from the neutral pions is:

\begin{equation}\label{Fgam}
q_{\gamma}(E_{\gamma})=2\int^{\infty}_{E_{\pi}^{\rm min}}
\frac{q_{\pi}(E_{\pi})}{\sqrt{E_{\pi}^{2}- m_{\pi}^{2} c^4}}
\;dE_{\pi},
\end{equation}
where
\begin{equation}\label{Epi}
E_{\pi}^{\rm min}(E_{\gamma})=E_{\gamma}+\frac{m_{\pi}^{2}
c^4}{4E_{\gamma}},
\end{equation}
and
\begin{eqnarray}
q_{\pi}(E_{\pi})&=&\int\delta (E_{\pi}-f_{\pi}(E-m_p c^2))
\sigma_{pp}(E)J_p(E)\;dE\nonumber\\
&=&\frac{1}{f_{\pi}}\sigma_{pp}\left(m_pc^2+\frac{E_{\pi}}{f_{\pi}}\right)
\,J_p\left(m_pc^2+\frac{E_{\pi}}{f_{\pi}}\right).
\end{eqnarray}
Here $f_{\pi}$ is the fraction of the kinetic energy of the proton that goes into the neutral pion (and then to gamma-rays). Then, using expression (\ref{J_p}) we obtain:
\begin{equation}
q_{\gamma}(E_{\gamma})\simeq \frac{J_{\rm
max}}{e\,\pi}\frac{\sigma_{pp}(E_p)}{f_{\pi}\,
E_p}\delta(E_{\gamma}-f_{\pi}E_p)\,{\rm
\frac{ph}{atom\,eV\,s\,sr}}.
\end{equation}
The intensity of the gamma-ray emission generated in the disk through $\pi^0$-decays results:
\begin{equation}
I_{\gamma}^o(E_{\gamma})= \int_0^{2h}
q_{\gamma}(E_{\gamma}){E_{\gamma}}^2 n_p\,
e^{-z/\lambda_{pp}}\;dz.
\end{equation}
The energy released per unit of time in the disk by these photons is 
\begin{equation}
L^{\pi^0}(E_{\gamma})=2\,f_{\pi}\,\dot{E}_{\rm inj}
(1-e^{-2h/\lambda_{pp}}),
\end{equation}
and the gamma-ray luminosity emerging from the other side of the disk is:
\begin{equation}
	L^{\pi^0}_{\rm esc}(E_{\gamma})=L^{\pi^0}(E_{\gamma}) e^{-2h\sigma_{\gamma p}n_p}. \label{L_esc}
\end{equation}
Here, $\sigma_{\gamma p}\sim 10^{-26}$ cm$^2$. The gamma-rays that are absorbed can initiate electromagnetic cascades inside the disk.

\section{Time-dependent estimates}                  

\begin{figure} 
\vskip 2.3in
\includegraphics{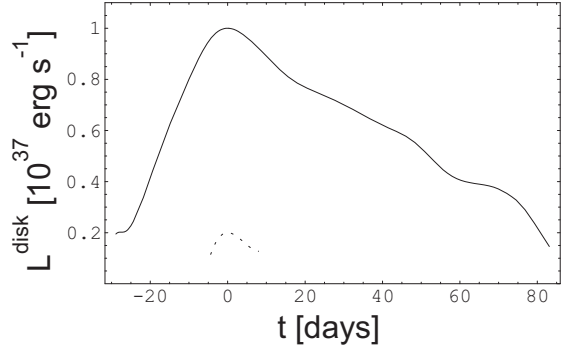}
\caption[]{Time evolution of the X-ray luminosity of the disk for a major ($L_{37}^{\rm peak}=1$) and a normal outburst ($L_{37}^{\rm peak}=0.2$, dashed curve). $t=0$ corresponds to the periastron passage. }
\label{fig3}
\end{figure}

In order to calculate the evolution of all relevant parameters during both a major and a normal outburst, we have adopted the X-ray light curves observed for the Be/X-ray transient 2S 1417-624 (Finger et al. 1996b), with a normalization\footnote{The distance to this source is not well-established} such that at the peak of the major outburst, which is coincident with the periastron passage, the luminosity is $L_{37}=1$. The light curves are shown in Fig. \ref{fig3}. 

\begin{figure} 
\vskip 3.7in
\includegraphics{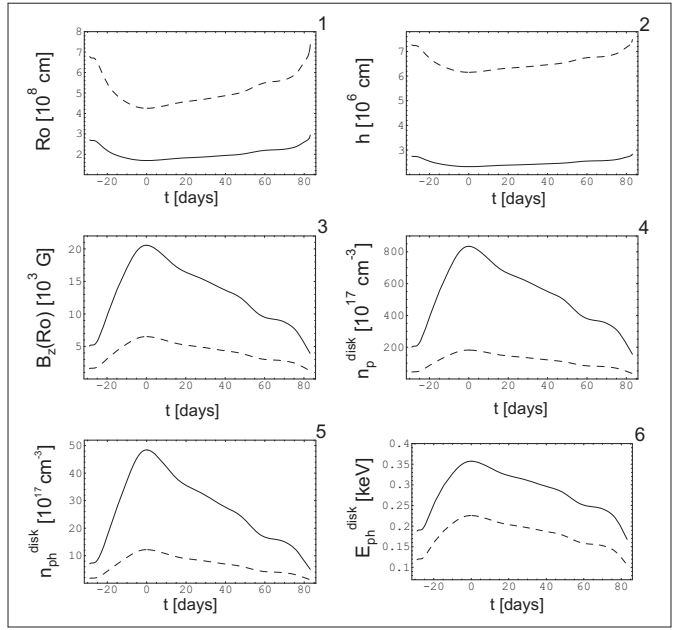}
\caption[]{Time-evolution of disk parameters during a major outburst for models with $\eta=1$ (dashed curve) and $\eta=0.2$ (solid curve). $t=0$ corresponds to the periastron passage. 1. Inner radius of the accretion disk ($R_0$). 2. Half-height of the accretion disk at $R_0$. 3. Magnetic field inside the disk at $R_0$. 4. Particle density of the disk at $R_0$. 5. Photon density in the disk at $R_0$. 6. Average energy of the disk photons at $R_0$.}
\label{fig4}
\end{figure}

\begin{figure} 
\vskip 3.7in
\includegraphics{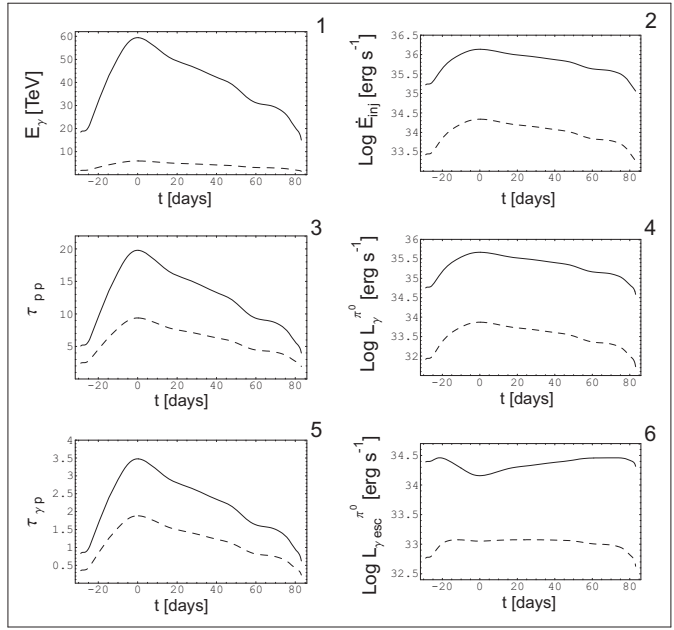}
\caption[]{Time-evolution of different parameters related to the high-energy emission during a major outburst for models with $\eta=1$ (dashed curve) and $\eta=0.2$ (solid curve). $t=0$ corresponds to the periastron passage. 1. Energy of the first-generation photons from $\pi^0$-decays. 2. Power injected into the disk in the form of relativistic protons. 3. Opacity to proton propagation in the disk. 4. Gamma-ray power from $\pi^0$-decays produced inside the disk. 5. Opacity to TeV gamma-ray propagation in the disk. 6. High-energy gamma-ray luminosity that emerges from the other side of the disk.}
\label{fig5}
\end{figure}

\begin{figure} 
\vskip 3.7in
\includegraphics{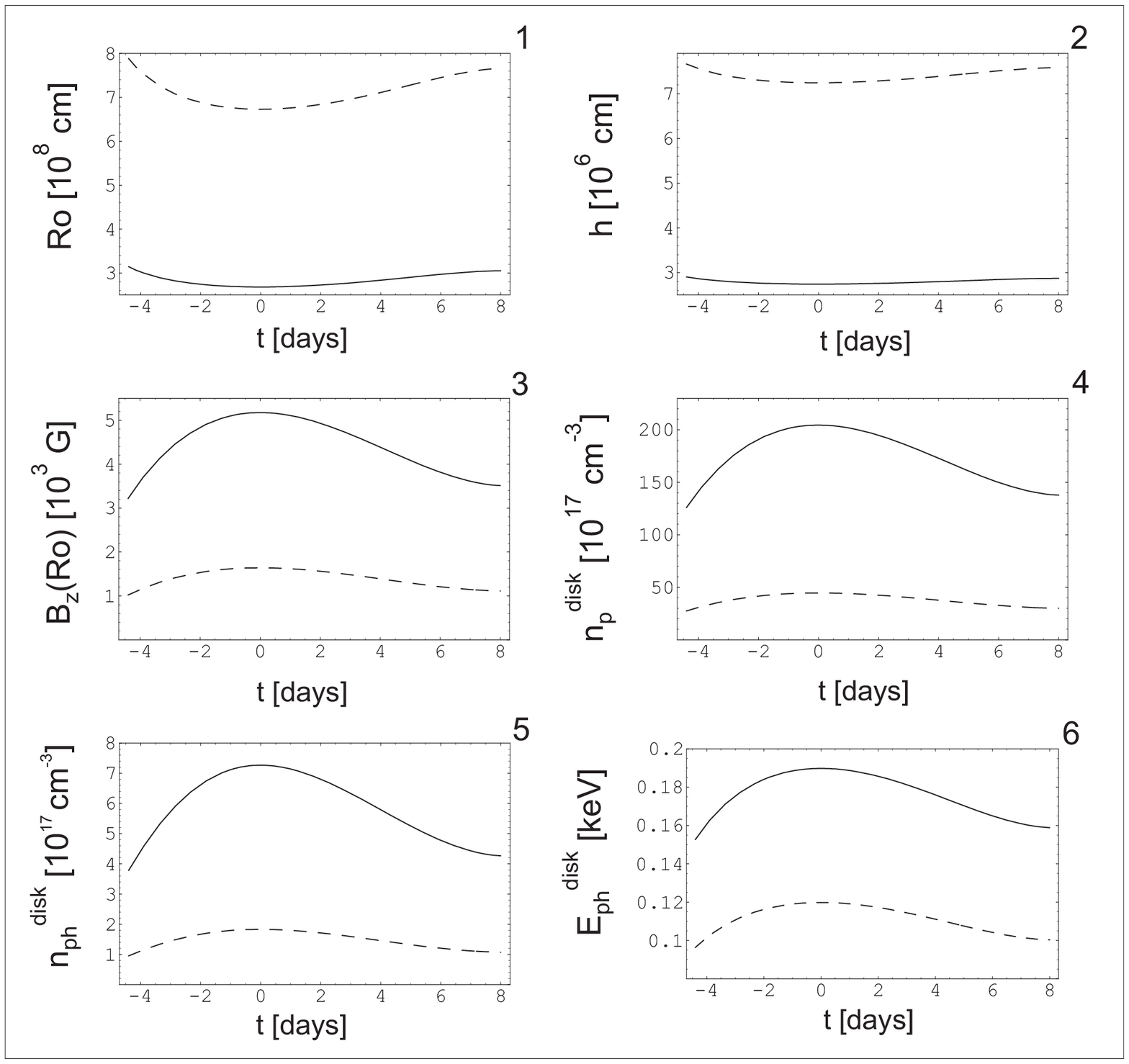}
\caption[]{Idem Fig. \ref{fig4}, but for a normal outburst.}
\label{fig6}
\end{figure}

\begin{figure} 
\vskip 3.7in
\includegraphics{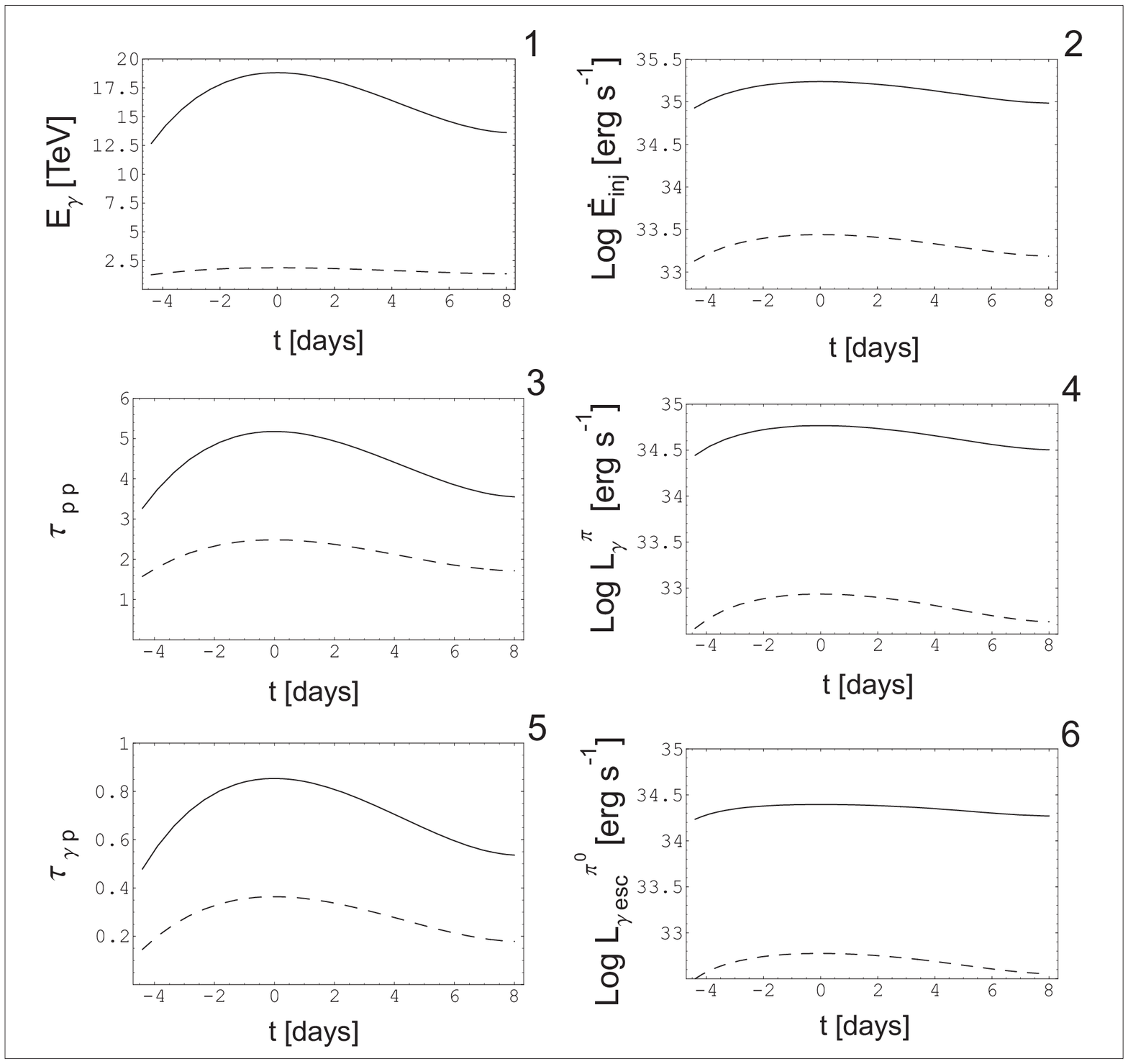}
\caption[]{Idem Fig. \ref{fig5}, but for a normal outburst.}
\label{fig7}
\end{figure}

In Figs. \ref{fig4}-\ref{fig7} we show the evolution of different parameters that characterizes the system and the high energy emission for the cases of an unscreened disk ($\eta=1$) and a partially screened disk ($\eta=0.2$).  The relation of the screening factor $\eta$ and the inner radius of the accretion disk is given by (Wang 1996): 
\begin{equation}
R_0=1.35 \gamma_0^{2/7} \eta^{4/7} r_{\rm A},	
\end{equation}
where $\gamma_0=-B_{\phi_0}/B_{z_0}\sim 1$ is the magnetic pitch angle. 

In the figures we show, for each case, the evolution of the inner radius of the accretion disk, the half-height of the disk, the $z$-component of the magnetic field, the particle density inside the disk, the photon density, the average energy of the photons, the energy of the first-generation gamma-rays inside the disk, the injected energy into the disk in the form of relativistic particles, the opacity of the disk to proton propagation, the gamma-ray luminosity produced inside the disk as a consequence of the $\pi^0$-decays, the opacity of the disk to these high-energy gamma-rays, and the luminosity of the gamma-rays that finally emerge from the other side of the disk. All values are given at $R_0$.  

\begin{figure} 
\vskip 3.2in
\includegraphics{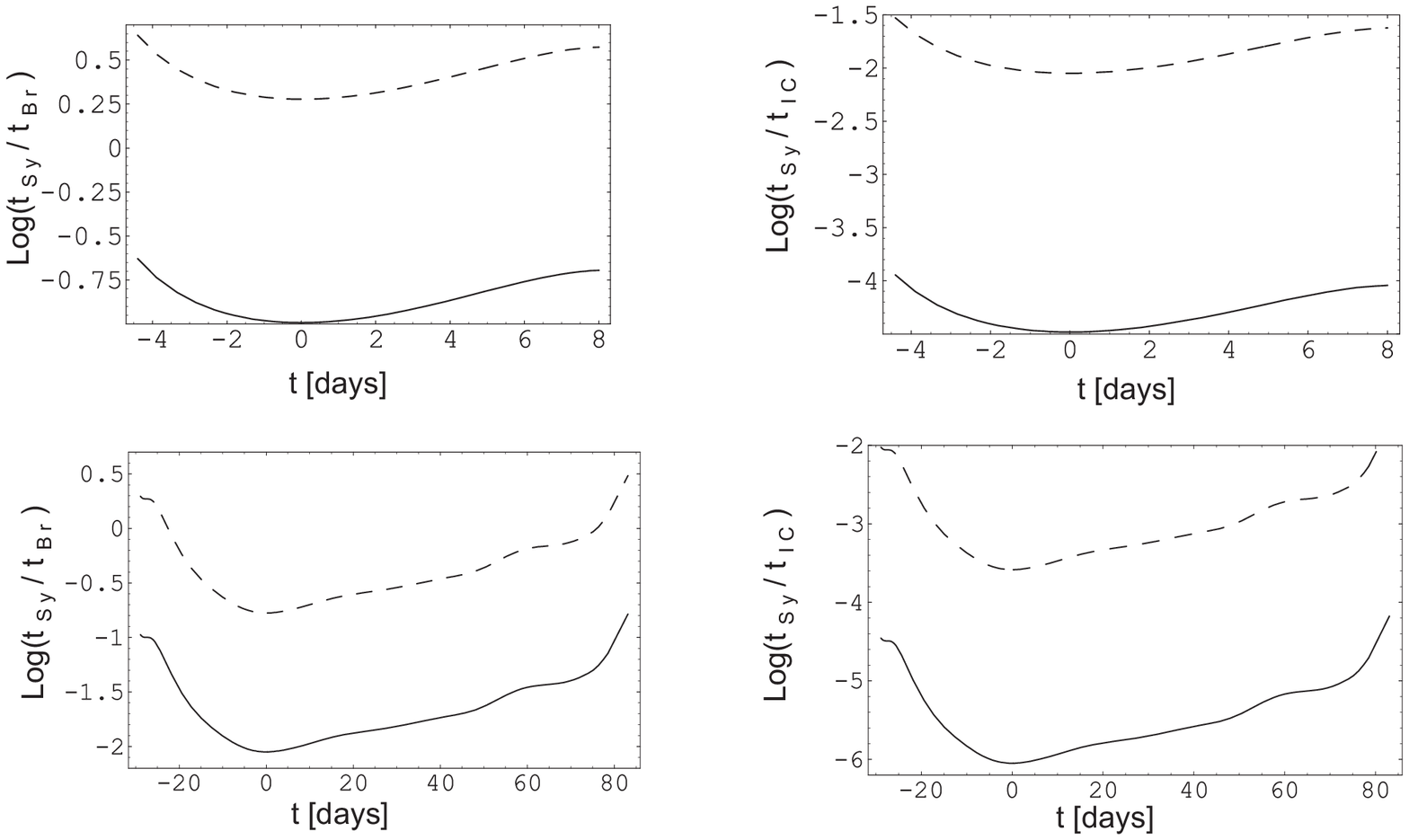}
\caption[]{Evolution of the different ratios of cooling times for secondary pairs in an average electromagnetic cascade developed in the accretion disk of a systems with screening factor $\eta=0.2$ (dashed) and $\eta=1$ (solid) during normal (upper panels) and major (lower panels) outbursts.}
\label{fig8}
\end{figure}

As we have already mentioned, those gamma-rays that are absorbed in the disk can initiate an electromagnetic cascade if the secondary pairs are energetic enough as to generate a second generation of gamma-rays. In Figure \ref{fig8} we show the evolution of the different ratios of cooling times for secondary pairs in the disk. During major outbursts in models with $\eta=0.2$ three generations of pairs are created. The first generation cools mainly by synchrotron radiation, whereas in the other generations the cooling is dominated by relativistic Bremsstrahlung. The GeV gamma-rays produced by the last generation of pairs escape from the disk but not necessarily from the system.  Inverse Compton losses, which were not dominant inside the disk, are now responsible for further cascades in the photosphere (see next section). In  models of major outbursts with $\eta=1$ only two generations of pairs are produced close to the periastron, and just one when the disk is far from it. These pairs cool by synchrotron radiation mainly, although Bremsstrahlung takes over far in the extremes of the light curve. During minor outbursts cascades cannot be sustained: TeV gamma-rays from $\pi$-decays and MeV-GeV photons from synchrotron and Bremsstrahlung cooling of pairs originated in the $\pi^{\pm}$-decays are directly injected  through the disk into the photosphere.       

\section{Absorption in the photosphere}

\begin{figure} 
\vskip 1.9in
\includegraphics{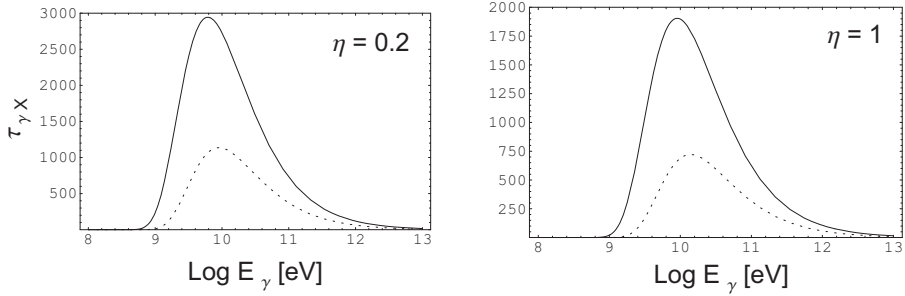}
\caption[]{Opacity to gamma-ray propagation in the photosphere at the periastron passage for both major (solid) and normal (dashed) outbursts and disks with $\eta=0.2$ (left) and $\eta=1$ (right).}
\label{fig9}
\end{figure}

\begin{figure} 
\vskip 1.9in
\includegraphics{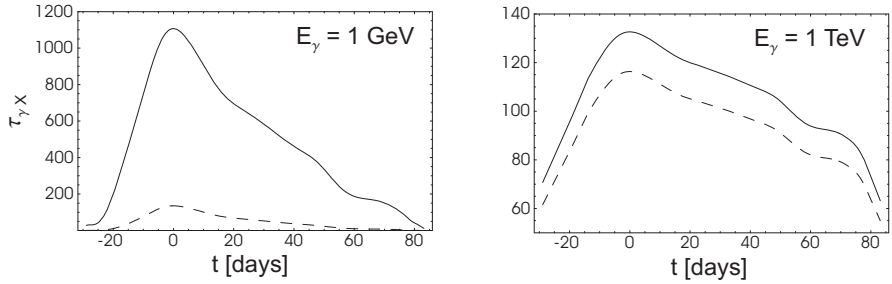}
\caption[]{Time evolution of the opacity to gamma-ray propagation in the photosphere for a major outbursts in disks with $\eta=0.2$ (solid curve) and $\eta=1$ (dashed curve) at two different energies.}
\label{fig10}
\end{figure}

\begin{figure} 
\vskip 1.9in
\includegraphics{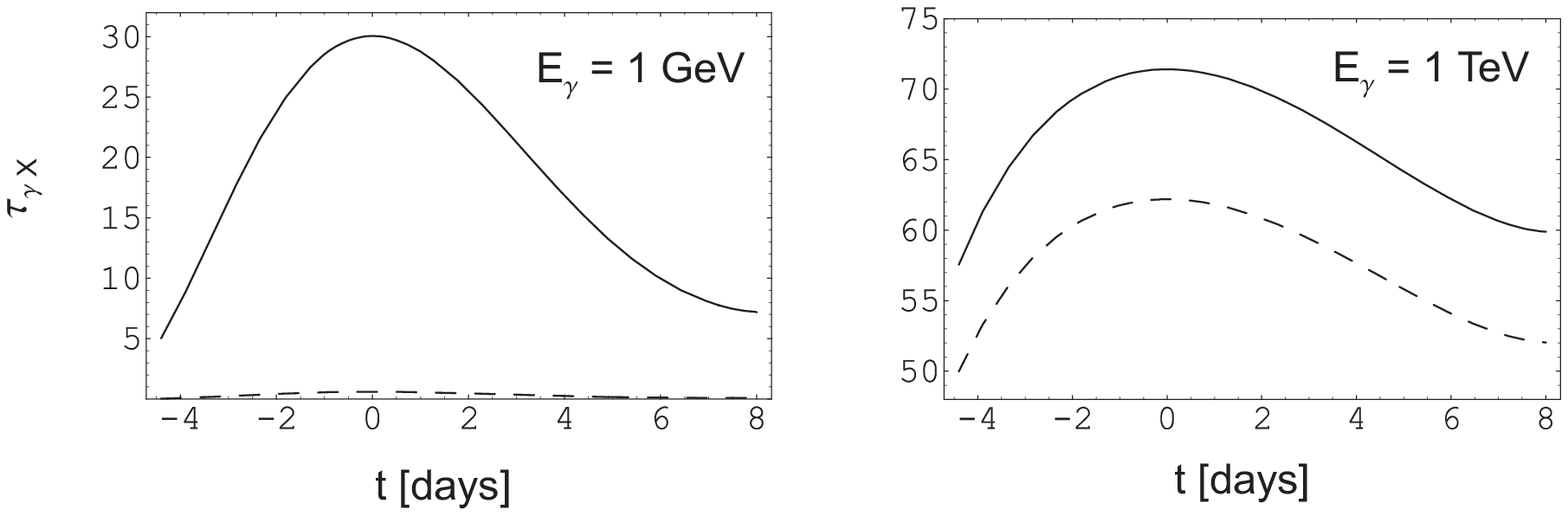}
\caption[]{Idem Fig. \ref{fig10} for a normal outburst.}
\label{fig11}
\end{figure}

The optical depth for a photon with energy $E_{\gamma}$ in the photosphere of the accreting neutron star is
 
\begin{equation}
\tau(R_o,\,E_{\gamma})=\int_{E_{\rm
min}(E_{\gamma})}^\infty\int_{R_o}^\infty\,n_{\rm ph}(E_{\rm
X},r)\sigma_{e^-e^+}(E_{\rm X},E_{\gamma})dr\,dE_{\rm X}, \label{tauXg}
\end{equation}
where $E_{\rm X}$ is the energy of the ambient thermal X-ray photons, $n_{\rm
ph}(E_{\rm X},r)$ is their density, and $\sigma_{e^-e^+}(E_{\rm X},E_{\gamma})$ is the photon-photon pair creation cross section given by:
\begin{equation}
\sigma_{e^+e^-}(E_{\rm X}, \;E_{\gamma})=\frac{\pi
r_0^2}{2}(1-\xi^2)\left[2\xi(\xi^2-
2)+(3-\xi^4)\ln\left(\frac{1+\xi}{1-\xi}\right) \right],
\end{equation}
where $r_0$ is the classical radius of the electron and 
\begin{equation}
\xi=\left[1-\frac{(m_e c^2)^2}{E_{\rm X} E_{\gamma}}\right]^{1/2}.
\end{equation}
In Eq. \ref{tauXg}, $E_{\rm min}$ is the threshold energy for pair creation in the X-ray field. This field can be considered as formed by two components, one from the disk and the other from the neutron star: $n_{\rm ph}=n_{\rm ph,1}+ n_{\rm ph,2}$. Here,
\begin{equation}
n_{\rm ph,1}(E_{\rm X},r)= \left(\frac{\pi B(E_{\rm
X})}{hc\,E_{\rm X}}\right)\frac{R_o^2}{r^2},
\end{equation}
with
\begin{equation}
B(E_X)= \frac{2 E_{\rm X}^3}{(hc)^2\,(e^{E_{\rm X}/kT_{\rm disk}}-1)}
\end{equation}
and $T_{\rm disk}=E_{\rm ph,disk}/k$, is the disk contribution. The emission from the heated polar caps of the star can be approximated by a Bremsstrahlung spectrum:
\begin{equation}
n_{\rm ph,2}(E_{\rm X},r)=\frac{L_o}{4\pi c\,r^2\,E_{\rm
X}^2\,e^{E_{\rm X}/kT}}\mbox{ for $E_{\rm X}\geq 2$ keV},
\end{equation}
where $L_o$ is the total luminosity in hard X-rays and $kT\sim 100$ keV.

In Figure \ref{fig9} we show the opacity to gamma-ray propagation at the periastron passage as a function of the gamma-ray energy for major and normal outbursts in both models with $\eta=0.2$ and $\eta=1$. We see that gamma-rays with energies between $\sim1$ GeV and $\sim1$ TeV energies are absorbed. These gamma-rays initiate inverse Compton cascades that degrade their energy. The final spectrum will be soft with an index of $\sim2.5$ and a high-energy cutoff around $\sim 2 m_e^2 c^4/ \left\langle E_{\rm X} \right\rangle\approx 250 (\left\langle E_{\rm X} \right\rangle/1\;{\rm keV})^{-1}$ MeV (Aharonian et al. 1985, Aharonian \& Plyasheshnikov 2003). In Figures \ref{fig10} and \ref{fig11} we show the time evolution of the opacity at 1 GeV and 1 TeV for different models and outburst types. GeV gamma-rays can escape only from minor outbursts and no screening of the magnetic field in the disk. Photons with energies above 10 TeV and below 500 MeV always can escape from the photosphere.      

\section{Discussion}

The variability pattern of a given source at gamma rays will depend on both the physical parameters that characterize the particular system and the energy band of the observations. For instance, almost all GeV sources will be strongly absorbed in their X-ray photosphere during major outbursts. In some normal outbursts, however, a flaring GeV event might be observable. At TeV energies the gamma rays are partially absorbed in the disks, especially during major outbursts (a fact already noticed by Cheng et al. 1991). The TeV flux can be detectable for sources in the inner spiral arms of the Galaxy, once again, mainly during normal outbursts. During major flaring X-ray events, the TeV gamma rays can be variable or not, depending on the particular combination of parameters. At MeV energies the sources should be detectable for a wide range of parameters, for both normal and major bursts. A correlation between the X-ray and the MeV gamma-ray flux should be observed in these systems, whereas an anti-correlation should be the case in GeV sources, due to the different absorptions.

Future detection of Be/X-ray binaries by GLAST and Cherenkov telescopes might be favored by the observation of the candidates during normal X-ray outbursts.     


\acknowledgements

We have benefited from illuminating discussions with Felix Aharonian, Luis Anchordoqui, K.S. Cheng, and Diego Torres.
This work has been supported by Fundaci\'on Antorchas, ANPCyT, and CONICET (PICT 0438/98). GER thanks 
the SECyT (Argentina) for a Houssay Prize.

\end{article}
\end{document}